\documentclass[aps,prl,twocolumn,amsmath,amssymb,nofootinbib,superscriptaddress,floatfix,reprint,longbibliography]{revtex4-1}
\usepackage[dvips]{graphicx}
\usepackage{latexsym}
\usepackage{amsmath}
\usepackage{amsfonts}
\usepackage{amssymb}
\usepackage{color}
\usepackage{txfonts}
\usepackage{float}
\usepackage{url}
\usepackage[colorlinks=true, urlcolor=blue, linkcolor=blue, citecolor=blue]{hyperref}
\usepackage{ulem}
\usepackage{tikz-feynman}
\usepackage{tikz,tikz-feynhand} 
\usepackage{graphicx}
\usepackage{extpfeil}
\usepackage{subfigure}
\usepackage{physics}
\begin{document}
	\newcommand{\fig}[2]{\includegraphics[width=#1]{#2}}
	\newcommand{\la}{{\langle}}
	\newcommand{\ra}{{\rangle}}
	\newcommand{\dg}{{\dagger}}
	\newcommand{\upa}{{\uparrow}}
	\newcommand{\dna}{{\downarrow}}
	\newcommand{\ab}{{\alpha\beta}}
	\newcommand{\ias}{{i\alpha\sigma}}
	\newcommand{\ibs}{{i\beta\sigma}}
	\newcommand{\hH}{\hat{H}}
	\newcommand{\hn}{\hat{n}}
	\newcommand{\hc}{{\hat{\chi}}}
	\newcommand{\hU}{{\hat{U}}}
	\newcommand{\hV}{{\hat{V}}}
	\newcommand{\br}{{\bf r}}
	\newcommand{\bk}{{{\bf k}}}
	\newcommand{\bq}{{{\bf q}}}
	\def\gsim{~\rlap{$>$}{\lower 1.0ex\hbox{$\sim$}}}
	\setlength{\unitlength}{1mm}
	\newcommand{{\vhf}}{$\chi^\text{v}_f$}
	\newcommand{{\vhd}}{$\chi^\text{v}_d$}
	\newcommand{{\vpd}}{$\Delta^\text{v}_d$}
	\newcommand{{\ved}}{$\epsilon^\text{v}_d$}
	\newcommand{{\vved}}{$\varepsilon^\text{v}_d$}
	\newcommand{\pprl}{Phys. Rev. Lett. \ }
	\newcommand{\pprb}{Phys. Rev. {B}}

\title {High Temperature Superconductivity in La$_3$Ni$_2$O$_7$}
\author{Kun Jiang}
\affiliation{Beijing National Laboratory for Condensed Matter Physics and Institute of Physics,
	Chinese Academy of Sciences, Beijing 100190, China}
\affiliation{School of Physical Sciences, University of Chinese Academy of Sciences, Beijing 100190, China}

\author{Ziqiang Wang}
\affiliation{Department of Physics, Boston College, Chestnut Hill, MA 02467, USA}

\author{Fuchun Zhang}
\affiliation{Kavli Institute of Theoretical Sciences, University of Chinese Academy of Sciences,
	Beijing, 100190, China}
\affiliation{Collaborative Innovation Center of Advanced Microstructures, Nanjing University, Nanjing 210093, China}

\date{\today}

\begin{abstract}
Motivated by the recent discovery of high-temperature superconductivity in bilayer La$_3$Ni$_2$O$_7$ under pressure, we study its electronic properties and superconductivity due to strong electron correlation. Using the inversion symmetry, we decouple the low-energy electronic structure into block-diagonal symmetric and antisymmetric sectors. We find that the antisymmetric sector can be reduced to a one-band system near half filling, while the symmetric bands occupied by about two electrons are heavily overdoped individually. Using the strong coupling mean field theory, we obtain strong superconducting pairing with $B_{1g}$ symmetry in the antisymmetric sector. We propose that due to the spin-orbital exchange coupling between the two sectors, $B_{1g}$ pairing is induced in the symmetric bands, which in-turn boosts the pairing gap in the antisymmetric band and enhances the high-temperature superconductivity with a congruent $d$-wave symmetry in pressurized La$_3$Ni$_2$O$_7$.
\end{abstract}
\maketitle

The discovery of high-temperature (high-T$_c$) superconductivity in the cuprates \cite{bednorz}, whose transition temperatures greatly exceed conventional superconductors, encourages exploring none-copper-based 
high-T$_c$ superconductors both experimentally and theoretically \cite{doping_mott,keimer_review,sc_wtcu, iron1,iron2,iron_review,hujp_PhysRevX.5.041012}. Among this exploration, it was theoretically proposed that the nickelates could be a counterpart of the cuprates \cite{rice_PhysRevB.59.7901,pickett_PhysRevB.70.165109}. Owing to sustained efforts on the synthesis \cite{nickelates0,nickelates1,nickelates2,nickelates3,nickelates4,nickelates5}, superconductivity was finally found in the ``infinite-layer" nickelates (Sr,Nd) NiO$_2$ thin films \cite{lidanfeng,lidanfeng2,lidanfeng3}, opening the Nickel age of superconductivity \cite{norman}.

\begin{figure}
	\begin{center}
		\fig{3.4in}{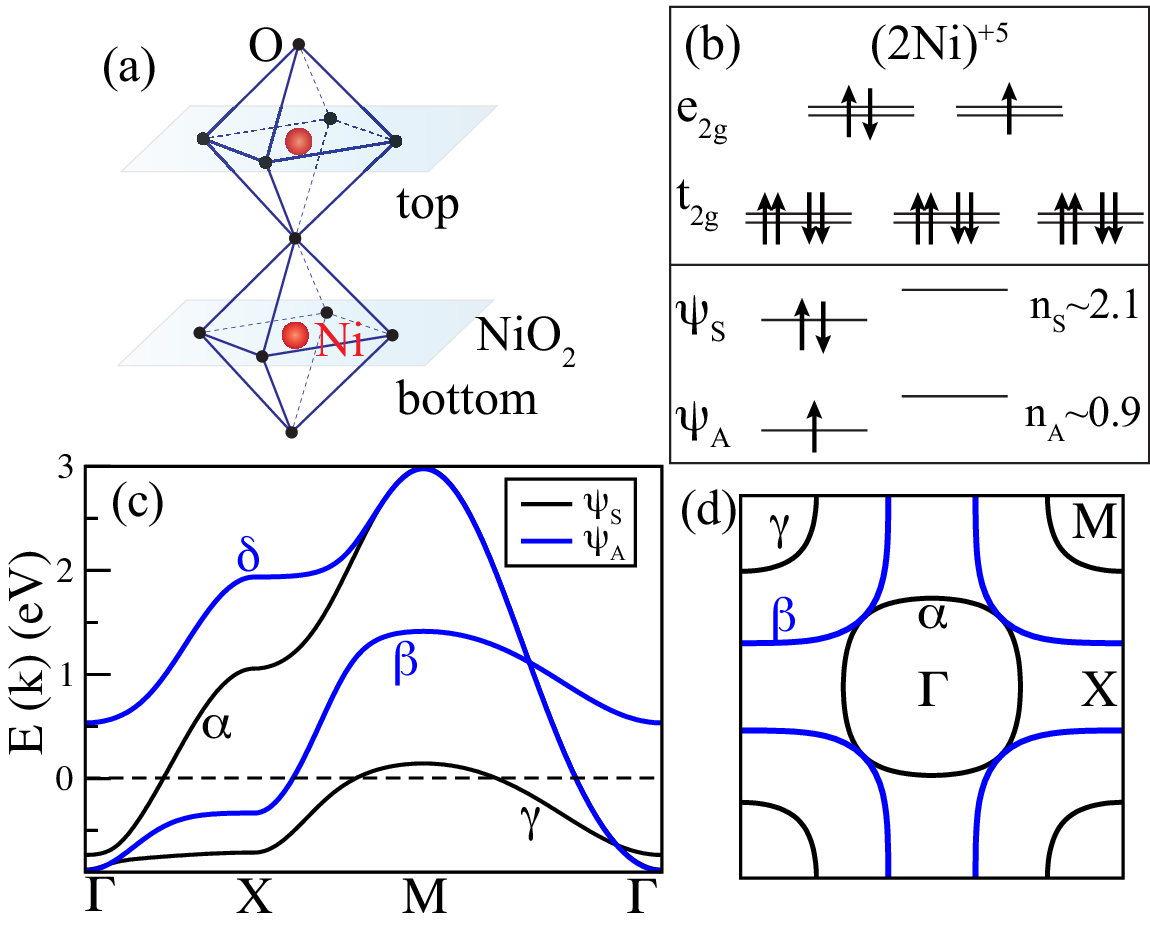}
		\caption{(a) The essential part of the LNO structure. The bilayer NiO$_2$ plane, labeled as the top and bottom layer, is formed by two corner-shared NiO$_6$ octahedrons. (b) Upper panel: the occupation configuration in the $(2Ni)^{+5}$ state ($Ni^{+2.5}$). Lower panel: the occupation configuration of the symmetric $\psi_{S}$ and antisymmetric $\psi_{A}$ orbitals. (c) The low-energy band structure decoupled into two $\psi_{S}$ bands (black) and two $\psi_{A}$ bands (blue). (d) The corresponding Fermi surfaces labeled by $\alpha$, $\beta$, and $\gamma$.
			\label{fig1}}
	\end{center}
\end{figure}

Recently, a new type of bulk nickelate La$_3$Ni$_2$O$_7$ (LNO) single crystal was successfully synthesized \cite{meng_wang}. A high-temperature superconducting transition $T_c\sim80$K under high-pressure was reported \cite{meng_wang,chengjg,yuanhq}. After its discovery, tremendous theoretical effort has been applied to this new material \cite{yaodx,dagotto1,wangqh,Kuroki,guyh,zhanggm,werner,leonov,wuwei,yangyf1,luy,yangf,wucj,dagotto2,zhangyh,siqm,sugang,yangyf2}.
Similar to the bilayer cuprates, the essential part of LNO superconductor is the bilayer NiO$_2 $ block \cite{meng_wang}, as illustrated in Fig.\ref{fig1} (a).  We label them as the top and bottom layer.
Around each Ni site, six oxygen atoms form a standard octahedron. The two nearest neighbor octahedrons between the two layers are corner shared by one apical oxygen. 
The LNO at ambient pressure is in its $Amam$ phase with the two octahedrons tilted. The phase evolves into the high symmetry structure $Fmmm$ phase under high pressure. The two octahedrons line up and superconductivity emerges around 14 Gpa. The octahedra crystal fields split the Ni 3d orbitals into $t_{2g}$ and $e_{g}$ complex, as shown in Fig. \ref{fig1}(b). 

Counting the chemical valence in LNO, $Ni$ is in the  $(2Ni)^{5+}$ state ($Ni^{2.5+}$ per-site). Notice that $Ni$ is normally in its $Ni^{2+}$ or $Ni^{1+}$ state, such that further hole doping always add holes into the oxygen \cite{fujimori_PhysRevB.30.957,sawatzky_PhysRevB.45.1612,sawatzky_PhysRevLett.62.221,shin_PhysRevLett.100.206401}. Therefore, the low-energy states of LNO are formed by the mixing $Ni-e_g$ and $O-p$ states, similar to the Zhang-Rice singlet in hole doped cuprates \cite{zhang_rice_PhysRevB.37.3759}. To simplify the discussion, we will continue to use $Ni$ 3$d$ states for convenience.
As shown in Fig, \ref{fig1}(b), the $(2Ni)^{5+}$ has fully occupied $t_{2g}$ orbitals and the $e_g$ orbitals host three electrons. In the following discussion, we label the $d_{x^2-y^2}$ and $d_{z^2}$ orbital as $d^{x}$ and $d^{z}$, and the top and bottom layers as $t$ and $b$.

Focusing on the partially occupied $e_g$ orbitals and utilizing the results of density functional theory (DFT) calculations, the tight-binding (TB) model for LNO can be derived \cite{yaodx,guyh} in the basis $(d_{tk}^{x},d_{tk}^{z},d_{bk}^{x},d_{bk}^{z})$ as
\begin{align}
H & ({\rm k})=\left(\begin{array}{cc}
H_{t}({\rm k}) & H_{\perp}({\rm k})\\
H_{\perp}({\rm k}) & H_{b}({\rm k})
\end{array}\right),\nonumber \\
H_{b}({\rm k})=H_{t}({\rm k})= & \left(\begin{array}{cc}
T_{{\rm k}}^{x} & V_{{\rm k}}\\
V_{{\rm k}} & T_{{\rm k}}^{z}
\end{array}\right),\qquad H_{\perp}({\rm k})=\left(\begin{array}{cc}
t_{\bot}^{x} & V_{{\rm k}}^{\prime}\\
V_{{\rm k}}^{\prime} & t_{\bot}^{z}
\end{array}\right).\label{eq:tb}
\end{align}
Here,
$T_{{\rm k}}^{x/z}=t_{1}^{x/z}\gamma_k+t_{2}^{x/z}\alpha_k+\epsilon^{x/z}$, $V_{\rm{ k}}=t_{3}^{xz}\beta_k$, $V_{\rm{ k}}^{\prime}=t_{4}^{xz}\beta_k$ with $\gamma_k=2(\cos k_x+\cos k_y)$, $\alpha_k=4\cos k_x\cos k_y$, $\beta_k=2(\cos k_x-\cos k_y)$, and interalyer coupling $t_{\perp}^{x}=0.005$eV, $t_{\perp}^{z}=-0.635$eV.
The corresponding hopping parameters can be found in Ref.~\cite{yaodx} and in the supplemental materials (SMs). DFT calculations show that the interlayer coupling is significant in LNO, which is captured by the off-diagonal block $H_{\perp}({\rm k})$ of the TB Hamiltonian in Eq.~(\ref{eq:tb}). The LNO under pressure has an inversion symmetry about the shared apical oxygen. This means that $H(k)$ is block-diagonalized in the eigen basis of inversion  that exchanges the top and bottom layers,
\begin{eqnarray}
	\psi_{S}^{\eta}&=&(d_{t}^{\eta}+d_{b}^{\eta})/\sqrt{2} \\
	\psi_{A}^{\eta}&=&(d_{t}^{\eta}-d_{b}^{\eta})/\sqrt{2}
\end{eqnarray}
where $\eta=x,z$. It is easy to verify that $(\psi_{S}, \psi_{A})$ block diagonalizes $H ({\rm k})$ into 
\begin{eqnarray}
		H_{TB}(k)=\left(\begin{array}{cc} 
		H_S(k) & 0 \\
		0 & H_A(k)
	\end{array}\right).
\end{eqnarray}
The $H_{S/A}(k)$ takes the same form as $H_{t/b}({\rm k})$ in Eq.(\ref{eq:tb}) but with different hopping parameters, which are listed in the SM. 

The TB electronic structure is plotted in Fig.\ref{fig1}(c), which separates into two $\psi_{S}$ bands (black lines) and two $\psi_{A}$ bands (blue lines). 
There are three bands crossing Fermi level, which are labeled by $\alpha$, $\beta$, and $\gamma$ with an unoccupied $\delta$ band. The Fermi surfaces (FSs) consist of one electron pocket ($\alpha$) around the $\Gamma$ point and two hole pockets ($\beta$, $\gamma$) around the M points in the Brouillion zone as shown in Fig. \ref{fig1} (d). Using these symmetric and antisymmetric orbitals, we find the electron occupation number is quite  interesting: $\psi_{S}$ is occupied by close to two electrons and $\psi_{A}$ by close to one electron, as summarized in Fig.\ref{fig1}(b). More precisely, the occupation in the anti-symmetric $\beta$ band is around 0.91, while the symmetric $\gamma$ band and $\alpha$ band are occupied by 1.725 and 0.365 electrons, respectively. 

\begin{figure}
	\begin{center}
		\fig{3.4in}{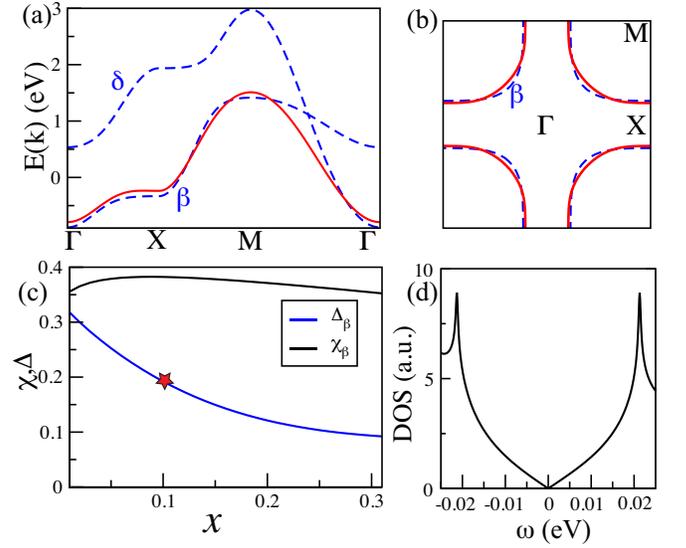}
		\caption{The band structure (a) and FS (b) for the antisymmetric $\psi_{A}$ bands plotted in blue dashed lines. The red solid lines correspond to those of the partially filled lower $\beta$ band fitted with a single-band $t-t'-t''$ model, where $t=0.288$eV, $t'=-0.0746$eV, $t''=0.04$eV. (c) The calculated mean-field order parameters $\chi_\beta$ and $\Delta_\beta$ as a function of hole doping $x$ for exchange coupling $J=0.12$ eV. The red star marks $x=0.1$ for the $\beta$ band, where the tunneling DOS is calculated and plotted in (d).
			\label{fig2}}
	\end{center}
\end{figure}

We start 
with the antisymmetric $\psi_{A}$ bands shown in Fig.~\ref{fig2}(a). Since the upper band is empty,  we can project out the upper $\delta$ band and focus on the $\beta$ band, which is close to half-filling. 
The orbital content of the $\beta$ band is dominated by the $d_{x^2-y^2}$ character in the DFT and the TB model. Hence, this is the band of the Zhang-Rice singlets \cite{zhang_rice_PhysRevB.37.3759}.
The dispersion of the $\beta$ band can be described by $t\gamma_k+t'\alpha_k+ t'' \gamma_{2k}$ with the effective nearest neighbor ($t$), next nearest neighbor ($t'$), and third neighbor hopping ($t''$), which is plotted (red line) in Fig. \ref{fig2}(a). The hopping parameters $(t,t',t'')$ are given in the SM. The corresponding $\beta$-FS is shown in Fig. \ref{fig2}(b).
Since the $\beta$ band is about 10\% hole doped away from half-filling, the effects of local correlations are strong and captured by the one band $t$-$J$ model \cite{doping_mott},
\begin{equation}
	H_{\beta}=\sum_{ij} t_{ij} \hat{P} \psi_{\beta,i\sigma}^\dagger \psi_{\beta,j\sigma} \hat{P}  + \sum_{\langle ij\rangle} J({\bf S}_{i}\cdot{\bf S}_j-\frac{1}{4}n_i n_j).
\end{equation}
Here $\hat{P}$ is the projection operator that removes double occupancy, $J$ is the superexchange interaction and the Einstein summation notation over repeated indices is used. 

The projection operator can be handled by writing
$\psi_{\beta,i\sigma}=b_i^\dagger f_{i\sigma}$, where $b_i$ is a slave-boson keeping track of empty sites and $f_{i\sigma}$ is a spin-1/2 fermion keeping track of singly-occupied sites. A physical constraint $b_i^\dagger b_i+ f_{i\sigma}^\dagger f_{i\sigma}=1$ is enforced here. 
Following the standard slave-boson mean-field theory \cite{doping_mott}, $H_{\beta}$ can be approximated by
\begin{eqnarray}
	H_\beta^{MF}&=&\sum_{ij} \sqrt{x_ix_j} t_{ij} f_{i\sigma}^\dagger f_{j\sigma}+\frac{J}{4}  \sum_{\langle ij\rangle}(|\chi_{ij}|^2+|\Delta_{ij}|^2) \\
 \nonumber
	&-&\frac{J}{4} \sum_{\langle ij\rangle}(\chi_{ij}^* f_{i\sigma}^\dagger f_{j\sigma} + \Delta_{ij}^* f_{i\sigma} f_{j\sigma'} \epsilon_{\sigma \sigma'}+h.c),
	\label{sb}
\end{eqnarray}
where $\epsilon_{\sigma \sigma'}$ is the antisymmetric tensor, $\chi_{ij}=\langle f_{i\sigma}^\dagger f_{j\sigma} \rangle$ and $\Delta_{ij}=\epsilon_{\sigma \sigma'} \langle f_{i\sigma} f_{j\sigma'} \rangle$ are the mean-field nearest neighbor bond and spin-singlet pairing order parameters. The bosons $b_i$ are condensed to expectation values $\sqrt{x_i}$, where $x_{i}$ is the local doping concentration at site $i$. 
Choosing the homogeneous solution with $\chi_{ij}=\chi$, $x_i=x$, we find that the $B_{1g}$ pairing ansatz, with $\Delta_{x}=\Delta$ and $\Delta_{y}=-\Delta$ for bonds along $x$ and $y$ directions, is the ground state as in the cuprates \cite{doping_mott,Anderson_2004}. 
The mean-field order parameters are self-consistently calculated and plotted in Fig.\ref{fig2}(c), as a function of the hole doping level $x$.
For the $\beta$ band filling around $x=0.1$ (indicated by the red star in Fig.\ref{fig2}(c)), we obtain $\chi=0.38$ and $\Delta=0.19$.  The calculated tunneling density of states (DOS) is shown in Fig.\ref{fig2}(d) at $x=0.1$, exhibiting a large pairing gap of $21.3$ meV.
Thus, independent of the precise value of $x$, the close to half-filled $\beta$ band plays the leading role in the high temperature superconductivity in LNO.

\begin{figure}
	\begin{center}
		\fig{3.4in}{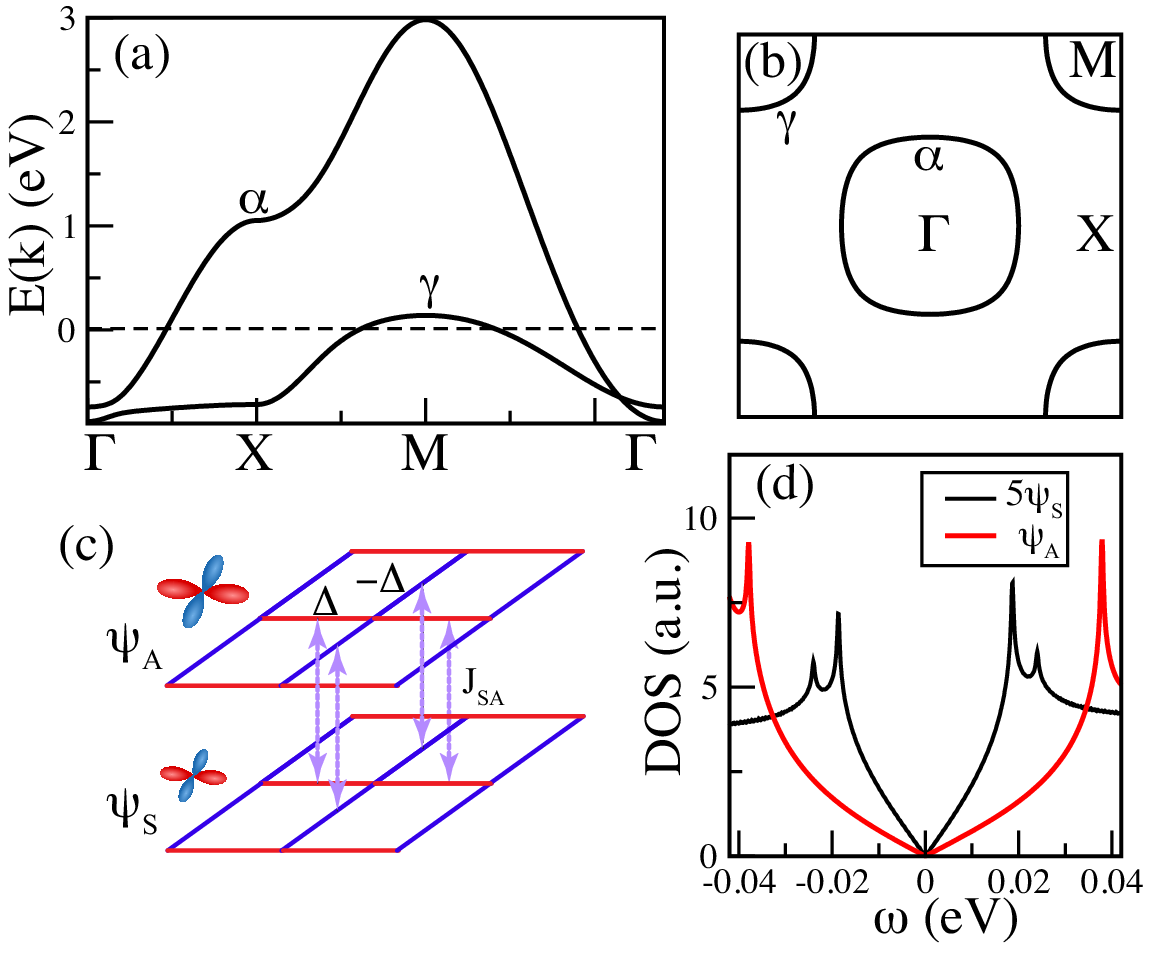}
		\caption{(a) The band structure (dashed blue) for the symmetric $\psi_{S}$ bands. (b) The corresponding FSs for $\psi_{S}$ bands (red). (c) The $B_{1g}$ superconductivity for the symmetric $\psi_{S}$ bands are induced by the antisymmetric $\psi_{A}$ through the spin-orbital coupling interaction $J_{SA}$, forming a congruent d-wave pairing state. The pairing order parameters are positive on the red bonds and negative on the blue bonds.
  (d) The tunneling density of states for the coupling model with $J_{SA}$=0.05eV. The DOSs for $\psi_S$ are magnified by a factor of 5 for visualization. 
			\label{fig3}}
	\end{center}
\end{figure}

Next, we consider the inversion symmetric $\psi_S$ bands and demonstrate that the high-$T_c$ superconductivity is further enhanced in a congruent $B_{1g}$ pairing state, such that the bilayer LNO can have a higher superconducting transition temperature $T_c$ than a single-layer cuprate such as La$_{2-x}$Sr$_x$CuO$_4$ as observed experimentally \cite{meng_wang,chengjg,yuanhq}. The dispersion of the two $\psi_S$ bands, with a filling fraction of
$n_S \sim 2.1$, and the corresponding FSs are plotted in Figs.\ref{fig3}(a) and (b).  As discussed above, there is one hole-like $\gamma$ FS corresponding to electron filling $n_\gamma=1.725$ centered around the $M$ point and one electron-like $\alpha$ FS with $n_\alpha=0.365$ around the $\Gamma$ point. This situation is similar to the iron-based superconductors and highly doped monolayer CuO$_2$ with the liberated $d_{z^2}$ orbital \cite{kun_PhysRevLett.121.227002,kun_PhysRevB.103.045108}. 
If we only consider these two $\psi_S$ bands, we find that the leading pairing channel is $A_{1g}$ with anti-phase $S_{\pm}$  order  parameters at $\alpha$ and $\gamma$ FSs at $n_S \sim 2.1$ filling.  However, the pairing order parameters obtained here are 10 times smaller than the $B_{1g}$ pairing  order parameter in the $\beta$ band. On the other hand, since the $\psi_S$ bands are highly overdoped with respect to half-filling in each band, the symmetric sector is far away from a doped Mott insulator and the effects of local correlation such as the band narrowing are relatively weak. As a result, the DOS of the whole system is dominated by the strongly renormalized antisymmetric $\beta$ band. Hence, it is important to consider the coupling between the symmetric and antisymmetric sectors for the superconducting state of LNO.

Microscopically, although the inversion symmetry decouples the $\psi_S$ and $\psi_A$ bands for single-particle excitations, the Coulomb interactions couple the two sectors, which is discussed in more detail in the SM. For our consideration, the most important symmetry allowed coupling is the spin-orbital exchange interaction \cite{kun_PhysRevB.103.045108,kun_PhysRevLett.121.227002},
\begin{eqnarray}
	H_{SAS}=J_{SA} (\hat{\Delta}_{Sx}^\dagger \hat{\Delta}_{A\beta}+ \hat{\Delta}_{Sz}^\dagger \hat{\Delta}_{A\beta}+h.c.), \label{sa}
\end{eqnarray}
which serves as an effective Josephson coupling between the pairing order parameters in the symmetric and antisymmetric sectors. 
Choosing a moderate $J_{SA}=0.05$ eV and ignoring the weak band renormalization of the overdoped symmetric sector, 
we calculate the pairing order parameters self-consistently by solving for the ground state of $H_\beta^{MF}+H_S^{MF}+H_{SAS}$ (see SM for more details). Intriguingly, the $B_{1g}$ superconducting state in the $\beta$ band, through the coupling $J_{SA}$, drives a congruent $B_{1g}$ pairing state in the symmetric $\alpha$ and $\gamma$ bands, as illustrated in Fig.\ref{fig3}(c). 
Specifically, all order parameters have a $d$-wave symmetry with amplitudes $(\Delta_{A,\beta}; \Delta_{S,x},\Delta_{S,z})=(0.24;0.02,0.04)$.
The calculated tunneling DOS plotted in Fig.\ref{fig3}(d) shows that the contribution from the $\beta$ band dominates and the spectral weight from the $\psi_S$ bands is magnified by a factor of 5 for visualization.
As clearly seen in Fig. \ref{fig3}(d), there are two energy gaps from the $\psi_S$ bands at $24.0$ meV and $18.7$ meV. Remarkably, the DOS reveals a large gap around $37.9$ meV from the $\beta$ band, which is significantly larger than that produced by the uncoupled $\beta$ band $t$-$J$ model shown in Fig.\ref{fig2}(d). We thus conclude that exchange coupling the strongly correlated $\beta$ band to the weakly correlated $\alpha$ and $\gamma$ bands with a large carrier density produces a congruent $d$-wave pairing state with boosted pairing energy gap and enhanced high-$T_c$ superconductivity, which can be a novelty of LNO under pressure.


In summary, we have taken the viewpoint of strong interlayer hybridization and classified single electron state as symmetric ($k_z=0$) or antisymmetric ($k_z=\pi$) linear combination of the state on top and bottom layers. We then introduce a large “on-site” Coulomb repulsion in the single electron Hamiltonian for the symmetric and antisymmetric states, here a “site” means a molecule site consisting of two crystal sites with one in the top and one in the bottom layer. The Hamiltonian for the antisymmetric states describes a near half-filled hole doped $t$-$t’$-$t''$-$J$ model on a square lattice for predominantly Ni-3$d_{x^2-y^2}$ orbitals, which gives $d$-wave superconductivity, similar to superconductivity in cuprates. The Hamiltonian for symmetric states describes a near full-filled predominantly 3$d_{z^2}$ orbital band and a lightly filled predominantly 3$d_{x^2-y^2}$ orbital band, which do not appear to play dominant roles in superconductivity on their own. We argued by explicit calculations that the spin-orbital exchange coupling between the symmetric and antisymmetric sectors can drive a congruent $d$-wave pairing state with significantly boosted superconducting energy gap and thus enhanced transition temperature $T_c$, beyond those of the typical single-layer cuprates. This scenario agrees with the electron band calculations for the normal state \cite{meng_wang,yaodx,guyh}. Note that we have examined the Fermi surfaces based on the finite-U Gutzwiller approximation, and found that they remain approximately the same (see SM for more details). We thus expect that the angle resolved photoemission spectroscopy measurements would be consistent with the band calculations, which may support the present scenario. 
On the other hand, the inversion symmetry between the top and bottom NiO$_2$ layers with respect to the shared apical oxygen atoms in the middle of the bilayer is only present at high pressure. At ambient or low pressure, the symmetry description and hence the scenario presently in this paper does not apply to the system.

\subsection{Acknowledgement} K. J. and F. Z. acknowledge the support by the Ministry of Science and Technology  (Grant No. 2022YFA1403900), the National Natural Science Foundation of China (Grant No. NSFC-11888101, No. NSFC-12174428, No. NSFC-11920101005), the Strategic Priority Research Program of the Chinese Academy of Sciences (Grant No. XDB28000000, XDB33000000), the New Cornerstone Investigator Program, and the Chinese Academy of Sciences through the Youth Innovation Promotion Association (Grant No. 2022YSBR-048). Z.W. is supported by the U.S. Department of Energy, Basic Energy Sciences Grant No. DE-FG02-99ER45747.

\bibliography{reference}

\clearpage
\onecolumngrid
\begin{center}
\textbf{\large Supplemental Material: High Temperature Superconductivity in La$_3$Ni$_2$O$_7$}
\end{center}

\setcounter{equation}{0}
\setcounter{figure}{0}
\setcounter{table}{0}
\setcounter{page}{1}
\makeatletter
\renewcommand{\theequation}{S\arabic{equation}}
\renewcommand{\thefigure}{S\arabic{figure}}
\renewcommand{\bibnumfmt}[1]{[S#1]}
\renewcommand{\citenumfont}[1]{S#1}

\twocolumngrid
\subsection{Tight-binding parameters}
The tight-binding hopping parameters for the bilayer two-orbital model in Eq. 1 of the main text are obtained from Ref.\cite{yaodx,guyh}. Their parameters are listed in Tab.\ref{tab:hop1}.
\begin{table}
	\begin{tabular}{ccccc}
		\hline \hline 
		
		$t_{1}^{x}$ & $t_{1}^{z}$ & $t_{2}^{x}$ & $t_{2}^{z}$ & $t_{3}^{xz}$  \tabularnewline\hline 
		-0.483 & -0.110 & 0.069 & -0.017 & 0.239  \tabularnewline \hline 
		
		$t_{\bot}^{x}$ & $t_{\bot}^{z}$ & $t_{4}^{xz}$ & $\epsilon^{x}$ & $\epsilon^{z}$  \tabularnewline \hline 
		0.005 & -0.635 & -0.034 &  0.776 & 0.409 \tabularnewline
		
		\hline \hline
	\end{tabular}
	\caption{\label{tab:hop1}Tight-binding hopping  parameters in the main text Eq.~(1)  $H_(k)$ (unit here is eV).  $\epsilon^{x},\epsilon^z$ are site energies for Ni$-d_{x^2-y^2}, d_{3z^2-r^2}$ orbitals, respectively.    
	}
\end{table}

Using the $\psi_{S}$ and $\psi_{A}$ orbitals, the $H(k)$ is block-diagonized into
\begin{eqnarray}
	H_{TB}(k)=\left(\begin{array}{cc} 
		H_S(k) & 0 \\
		0 & H_A(k)
	\end{array}\right)  
\end{eqnarray}
And $H_S(k)$, $H_A(k)$ take the same structure of $H_t(k)$
\begin{eqnarray}
	H_{S/A}({\rm k})= & \left(\begin{array}{cc}
		T_{{\rm k}}^{x} & V_{{\rm k}}\\
		V_{{\rm k}} & T_{{\rm k}}^{z}
	\end{array}\right),
\end{eqnarray}
with
$T_{{\rm k}}^{x/z}=t_{1}^{x/z}\gamma_k+t_{2}^{x/z}\alpha_k+\epsilon^{x/z}, V_{\rm{ k}}=t_{3}^{xz}\beta_k$. Their parameters are listed in Tab.\ref{tab:hop2}.
\begin{table}
	\begin{tabular}{cccccccc}
		\hline \hline 
	sector	 &$t_{1}^{x}$ & $t_{1}^{z}$ & $t_{2}^{x}$ & $t_{2}^{z}$ & $t_{3}^{xz}$ & $\epsilon^{x}$ & $\epsilon^{z}$   \tabularnewline\hline 
	$\psi_{S}$&	-0.483 & -0.110 & 0.069 & -0.017 & 0.205  &  0.781  & -0.226 \tabularnewline \hline
	$\psi_{A}$&	-0.483 & -0.110 & 0.069 & -0.017 & 0.273 &  0.771 &  1.044  \tabularnewline 
		\hline \hline
	\end{tabular}
	\caption{\label{tab:hop2}Tight-binding hopping  parameters in $\psi_{S}$ and $\psi_{A}$ respectively. 
	}
\end{table}

For the fitting of the $\beta$ band, we use up to the third nearest neighbor hopping in the TB model on the square lattice $t\gamma_k+t'\alpha_k+ t'' \gamma_{2k}$. The parameters are $t=0.288$, $t'=-0.0746$, $t''=0.04$.

\subsection{The coupled Hamiltonian for $\psi_A$ and $\psi_S$}
As described in the main text, the coupled Hamiltonian for $\psi_A$ and $\psi_S$ can be written as $H_\beta^{MF}+H_S^{MF}+H_{SAS}$, where
$H_\beta^{MF}$ is the mean-field Hamiltonian defined in Eq.~(6) for the $\beta$ band. The $H_{SAS}$ is the coupling between two sectors in Eq.~(7).
The symmetric $\psi_S$ sector is described by a two-band model with the exchange interaction,
\begin{eqnarray}
   H_S=\sum_{ij} t_{ij}^{\eta,\eta'} \psi_{\eta,i\sigma}^\dagger \psi_{\eta',j\sigma}+\sum_{\langle ij\rangle} J({\bf S}_{i}\cdot{\bf S}_j-\frac{1}{4}n_i n_j)
\end{eqnarray}
Here $t_{ij}^{\eta,\eta'}$ are the hopping parameters in $H_S(k)$. The mean-field Hamiltonian $H_S^{MF}$ follows from decoupling the exchange interaction into the two-orbital pairings and bonds as in Eq.~(6). 
Notice that the band renormalization factors in $t_{ij}^{\eta,\eta'}$ are ignored because the $\psi_S$ bands are heavily doped away from half-filling individually.

\subsection{Hubbard Interactions in $\psi_{S}$ and $\psi_{A}$}
In this section, we discuss the interactions between $\psi_{S}$ and $\psi_{A}$.
Although the inversion symmetry blocks the hopping between $\psi_{S}$ and $\psi_{A}$, the Coulomb interactions between them are nonzero and take a multiorbital form.
More precisely, the local Hubbard interactions can be written as
\begin{eqnarray}
	\begin{split}
		H_I&= U\sum_{i,\eta} \hat{n}_{i,\eta \uparrow}\hat{n}_{i,\eta \downarrow}+U'\sum_{i,\eta\neq \eta^{\prime}}\hat{n}_{i,\eta}\hat{n}_{i,\eta^{\prime}}
		\\
		& -J_H\sum_{i,\eta\ne \eta^{\prime}} (\mathbf{S}_{i\eta} \cdot  \mathbf{S}_{i\eta'} 
		+d_{i,\eta\uparrow}^{\dagger}d_{i,\eta\downarrow}^{\dagger}d_{i,\eta^{\prime}\uparrow}d_{i,\eta^{\prime}\downarrow} )
	\end{split}
	\label{eq:HI}
\end{eqnarray}
where the $\eta$ is the orbital index.

Since the $\beta$ band mainly carries the $d_{x^2-y^2}$ character, we will simply use the $d_{x^2-y^2}$ for $\psi_\beta$.
Hence, the interaction between $d_{x}^{A}$ and $d_{x}^{S}$ coming from the intra-orbital $U\hat{n}_{i,\eta \uparrow}\hat{n}_{i,\eta \downarrow}$ takes the form 
\begin{equation}
	\begin{split}
		H_I^{A,S}&= U_0\sum_{i,\alpha} \hat{n}_{i,\uparrow}^{\alpha}\hat{n}_{i,\downarrow}^{\alpha}+U_v\sum_{i,\alpha\neq \alpha^{\prime}}\hat{n}_{i, \uparrow}^{\alpha}\hat{n}_{i,\downarrow} ^{\alpha^{\prime}}
		\\
		& -J\sum_{i,\alpha\ne \alpha^{\prime}} (d_{i,\uparrow}^{\alpha\dagger}d_{i,\downarrow}^{{\alpha^{\prime}}\dagger}d_{i,\uparrow}^{\alpha^{\prime}}d_{i,\downarrow} ^{\alpha}
		+d_{i,\uparrow}^{\alpha\dagger}d_{i,\downarrow}^{\alpha \dagger}d_{i,\uparrow} ^{\alpha^{\prime}} d_{i,\downarrow}^{\alpha^{\prime}} ),
	\end{split}
	\label{eq:Hsa}
\end{equation}
where $\alpha=S,A$ and $U_0=U_v=J=\frac{U}{2}$.

In the same spirit, we can decouple the inter-orbital interaction into a similar form. For example, the interaction between $d_{x}^{A}$ and $d_{z}^{S}$  comeing from $U' \hat{n}_{i,\eta}\hat{n}_{i,\eta^{\prime}}$ takes the form
\begin{equation}
	\begin{split}
		H_{I2}^{A,S}&= U_0\sum_{i,\alpha} \hat{n}_{i,x}^{\alpha}\hat{n}_{i,z}^{\alpha}+U_v\sum_{i,\alpha\neq \alpha^{\prime}}\hat{n}_{i, x}^{\alpha}\hat{n}_{i,z} ^{\alpha^{\prime}}
		\\
		& -J\sum_{i,\alpha\ne \alpha^{\prime}} (d_{i,x}^{\alpha\dagger}d_{i,z}^{{\alpha^{\prime}}\dagger}d_{i,x}^{\alpha^{\prime}}d_{i,z} ^{\alpha}
		+d_{i,x}^{\alpha\dagger}d_{i,z}^{\alpha \dagger}d_{i,x} ^{\alpha^{\prime}} d_{i,z}^{\alpha^{\prime}} )
	\end{split}
	\label{eq:Hsa2}
\end{equation}
with $\alpha=S,A$ and $U_0=U_v=J=\frac{U'}{2}$.
The Hund's rule interaction  $J_H \mathbf{S}_{i x} \cdot  \mathbf{S}_{iz}$ transforms into
\begin{eqnarray}
	&&H_{I3}^{A,S}=-J_0\sum_{i}  (\mathbf{S}_{i x}^S+ \mathbf{S}_{i x}^A)\cdot (\mathbf{S}_{i z}^S+ \mathbf{S}_{i z}^A) \\
 \nonumber
	&&-J_0 \sum_{i,\alpha\ne \alpha^{\prime}} (d_{i x \sigma}^{A\dagger} d_{i x \sigma'}^{S}+ d_{i x \sigma}^{S\dagger} d_{i x \sigma'}^{A}) \mathbf{\hat{S}}_{\sigma\sigma'}\cdot \mathbf{\hat{S}}_{\sigma'\sigma}  (d_{i z \sigma'}^{A\dagger} d_{i z \sigma}^{S}+ d_{i z \sigma'}^{S\dagger} d_{i z \sigma}^{A})
\label{eq:HI3}
\end{eqnarray}
with $J_0 =\frac{J_H}{2}$.
Collecting all the terms, the symmetry allowed local interactions are just the multi-orbital Hubbard model with the effective orbitals including with both the atomic orbitals and the molecular symmetric-antisymmetric sector index.
The inter-sector interactions are crucial and produce the inter-sector exchange interaction. 
As we discussed in previous works \cite{kun_PhysRevB.103.045108,kun_PhysRevLett.121.227002}, the inter-sector spin-orbital exchange interaction generates the effective Josephson coupling between the pairing order parameters, 
\begin{eqnarray}
	H_{SAS}=J_{SA} (\hat{\Delta}_{Sx}^\dagger \hat{\Delta}_{\beta}+ \hat{\Delta}_{Sz}^\dagger \hat{\Delta}_{\beta}+h.c.)
\end{eqnarray}

\subsection{Finite-U Gutzwiller approximation}
\begin{figure}
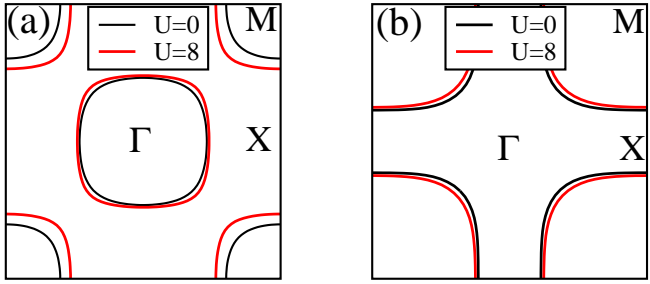

	\begin{center}
		\fig{3.4in}{fs_u8.eps}
		\caption{(a) FSs of $\psi_{S}$ at $U=0$ (black lines) and $U=8$ eV (red lines). (b) FSs of $\beta$ band at $U=0$ (black lines) and $U=8$ eV (red lines).
			\label{figfs}}
	\end{center}
\end{figure}

An important aspect of our theory is the doping concentration for the antisymmetric $\beta$ band and the symmetric $\alpha$ and $\beta$ bands. In the main text, we used the results of the DFT calculations, which are reproduced in the TB model. However, the strong local correlation can in principle generate inter-orbital and inter-sector charge transfer among the $\psi_A$ and $\psi_S$ bands by renormalizing the effective crystal fields. To this end, we carried out a finite-U multiorbital Gutzwiller approximation calculation \cite{kun_PhysRevB.103.045108,kun_PhysRevLett.121.227002}, including all four bands relevant for LNO. The results of the renormalized FSs are shown in Fig. \ref{figfs} for the Hubbard interaction $U=8$eV and Hund's coupling $J_H=$0.1$U$ and compared to the noninteracting case. Clearly, the correlation-induced charge transfer is weak as indicated by the small changes in the sizes of the FSs for correlation strength up to $U=8$eV, providing support for the results discussed in the main text.

\end{document}